\documentclass[prb,showpacs,superscriptaddress,twocolumn]{revtex4-2}
\usepackage[utf8]{inputenc}
\usepackage{amsmath,graphicx,hyperref}

\allowdisplaybreaks[1]
\setcitestyle{super}

\begin{document}

\title{Elastocaloric effect of the heavy-fermion system YbPtBi}

\author{Elena Gati}
\email{elena.gati@cpfs.mpg.de}
\affiliation{Max Planck Institute for Chemical Physics of Solids, 01187 Dresden, Germany}
\author{Burkhard Schmidt}
\affiliation{Max Planck Institute for Chemical Physics of Solids, 01187 Dresden, Germany}
\author{Sergey L. Bud'ko}
\affiliation{Ames National Laboratory, US Department of Energy, Iowa State University, Ames, Iowa 50011, USA}
\affiliation{Department of Physics and Astronomy, Iowa State University, Ames, Iowa 50011, USA}
\author{Andrew P. Mackenzie}
\affiliation{Max Planck Institute for Chemical Physics of Solids, 01187 Dresden, Germany}
\affiliation{Scottish Universities Physics Alliance, School of Physics and Astronomy, University of St Andrews, St Andrews, UK}
\author{Paul C. Canfield}
\affiliation{Ames National Laboratory, US Department of Energy, Iowa State University, Ames, Iowa 50011, USA}
\affiliation{Department of Physics and Astronomy, Iowa State University, Ames, Iowa 50011, USA}

\date{\today}

\begin{abstract}
YbPtBi is one of the heavy-fermion systems with largest Sommerfeld coefficient $\gamma$ and is thus classified as a `super'-heavy fermion material. In this work, we resolve the long-debated question about the hierarchy of relevant energy scales, such as crystal-electric field (CEF) levels, Kondo and magnetic ordering temperature, in YbPtBi. Through measurements of the a.c. elastocaloric effect and generic symmetry arguments, we identify an \textit{elastic level splitting} that is uniquely associated with the symmetry-allowed splitting of a quartet CEF level. This quartet, which we identify to be the first excited state at $\Delta/k_\text B\approx1.6\,\rm K$ above the doublet ground state at ambient pressure, is well below the Kondo temperature $T_\text K\approx10\,\rm K$. Thus, our analysis provides strong support for models that predict that the heavy electron mass is a result of an enhanced degeneracy of the CEF ground state, i.e., a quasi-sextet in YbPtBi. At the same time, our study shows the potential of the a.c. elastocaloric effect to control and quantify strain-induced changes of the CEF schemes, opening a different route to disentangle the CEF energy scales from other relevant energy scales in correlated quantum materials.
\end{abstract}

\pacs{xxx}

\maketitle

An enhanced effective electron mass $m^*$ is considered as a hallmark of a large class of strongly correlated metals. In such heavy-electron systems, many exotic quantum phenomena, including unconventional superconductivity, non Fermi-liquid behavior, quantum criticality and even topological semimetals~\cite{Stewart84,fulde:88,zwicknagl:92,thalmeier:04,Gegenwart08,Paschen21,Mun22} occur. Among the materials with the largest Sommerfeld coefficients $\gamma\propto m^*$ are the rare-earth based heavy fermion systems YbPtBi~\cite{Fisk91,Canfield91}, Yb$T_2$Zn$_{20}$ ($T=\rm Fe, Co$)~\cite{Torikachvili07} and PrAg$_2$In~\cite{Yatskar96}, also dubbed as `super' heavy electron systems~\cite{Canfield16,Tokiwa16}. In these cubic systems, $\gamma$ reaches record values as high as $10\,\rm J/(mol\,K)$. The Yb-variants have in common that they are characterized by small characteristic energy scales~\cite{Canfield94,Torikachvili07} of the order of $k_\text B\cdot1\ldots10\,\rm K$, including the Kondo scale $k_\text BT_\text K$ as well as excited crystal electric field (CEF) levels at $k_\text BT_\text{CEF}$. Correspondingly, it has been suggested that the hybridization of the conduction electrons~\cite{Stewart84} with a large number of degenerate CEF states is the source of the high electronic mass~\cite{Tsujii05,Torikachvili07,Shimura20}. However, given the multitude of small energy scales, a definite determination of the energy scales and thus a quantitative description of the unusually high $\gamma$ has proven difficult~\cite{Canfield94,Ueland15,Ivanshin09}.

\begin{figure}
\centering
\includegraphics[width=\columnwidth]{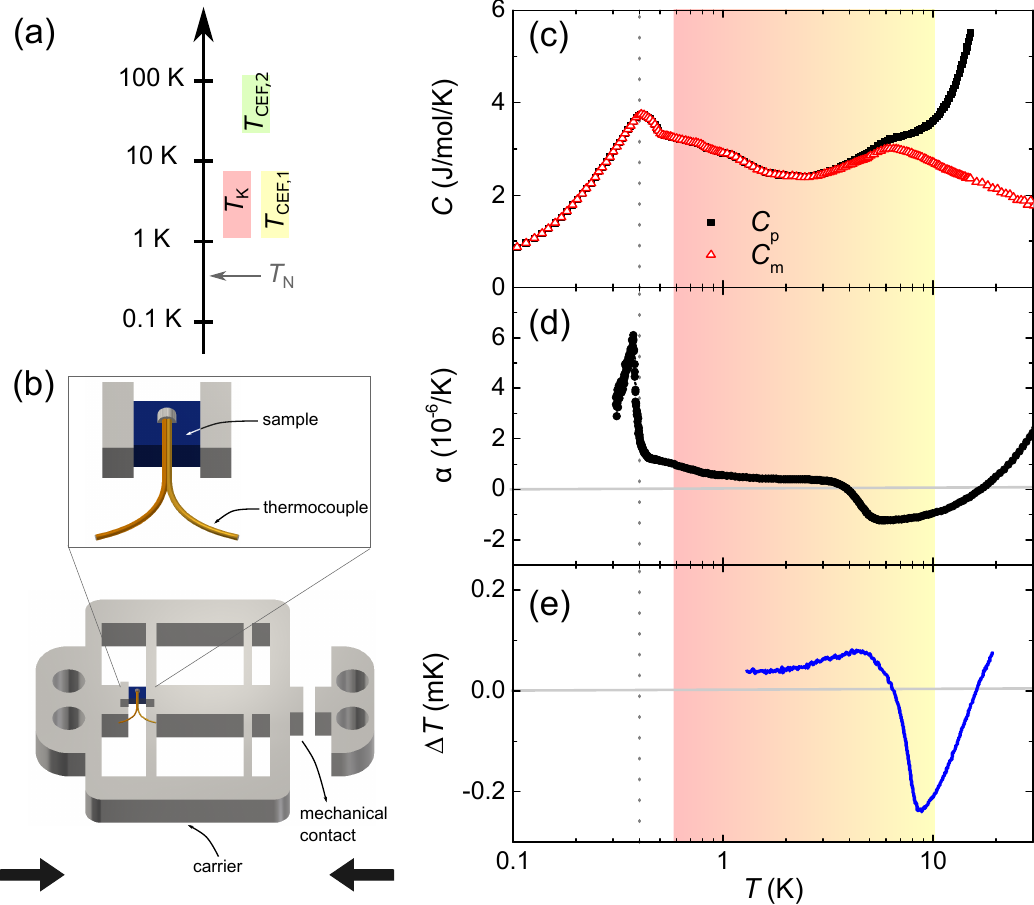} 
\caption{(a) Schematic representation of proposed characteristic energy scales of YbPtBi. $T_\textrm{N}$ indicates the position of magnetic ordering, $T_\textrm{K}$ the possible range of the Kondo crossover and $T_{\textrm{CEF},1}$ and $T_{\textrm{CEF},2}$ the possible temperature range of the first and second excited crystal-electric field levels. (b) Schematic of the setup to measure the elastocaloric effect \cite{Ikeda19}. A sample with attached thermocouple is placed across a gap on the sample carrier, which is screwed into the pressure cell (not shown). Different compressive uniaxial pressures can be exerted when the mechanical contact is closed by a uniaxial pressure cell. The direction of applied pressure is indicated by the big black arrows. The strain-induced temperature changes $\Delta T$ are recorded with the thermocouple. (c-e) Ambient-pressure properties of YbPtBi: total specific heat, $C_\textrm{p}$, and magnetic specific heat, $C_{\textrm{m}}$~\cite{Mun13}, (c) thermal expansion~\cite{Mun13}, $\alpha$, (d) and elastocaloric temperature amplitude, $\Delta T$ (e), from this work.}
\label{fig:overview}
\end{figure}

For YbPtBi, specifically, the following characteristic temperature scales~\cite{Fisk91,Canfield91,Canfield94,Robinson93,Robinson95,Ueland15} have been found so far (see Fig.~\ref{fig:overview}\,(a), (c) and (d)). Since the Yb$^{3+}$ Kramers ion resides on a site with cubic symmetry, the CEF is expected to split the $J=7/2$ multiplet into two doublets of type $\Gamma_6$ and $\Gamma_7$ and a $\Gamma_8$ quartet~\cite{Lea62}. The analyses of several experiments consistently find that the highest CEF excited state is a doublet (likely $\Gamma_6$) with $T_{\text{CEF},2}\sim60\ldots100\,\rm K$. Even though various studies \cite{Robinson95} favor $\Gamma_7$ to be the ground state and $\Gamma_8$ to be the first excited state with $T_{\text{CEF},1}\sim1\ldots10\,\rm K$, the reverse assignment with a $\Gamma_8$ ground state was also found to be compatible with a number of experimental results \cite{Robinson95}. In addition, symmetry-breaking distortions at low temperature, giving rise to additional level splittings, could not be conclusively ruled out so far~\cite{Robinson95,Robinson99,Martins95,Ueland15}. These uncertainties have not only hampered estimates for the absolute value of $T_{\text{CEF},1}$, but also of the second important temperature scale $T_\text K$~\cite{Canfield94} which is believed to be in the same energy range. Finally, for very low temperatures below $T_\text N\approx400\,\rm mK$, YbPtBi orders antiferromagnetically. This order is fragile~\cite{Mun13,Canfield16,Ueland14,Movshovich94} as it can be suppressed towards a quantum critical point by a small external magnetic field, and non Fermi-liquid behavior was discovered in the quantum critical region~\cite{Mun13}.

In this paper, we clarify the hierarchy of energy scales in YbPtBi, something only made possible through conducting thermodynamic measurements of the elastocaloric effect under well-controlled, symmetry-breaking, uniaxial pressure $p$. In contrast to magnetic field, which breaks time-reversal symmetry and thus affects all Kramers-degenerate states, the lattice strain $\epsilon$ associated with uniaxial pressure can only lift degeneracies stabilized by crystallographic symmetries. As we will show below, it is through this \textit{elastic level splitting} that the application of uniaxial pressure can be used to sensitively probe the single-ion physics \cite{Luethi05} associated with the $\Gamma_8$ state. As a result, we successfully disentangle the thermodynamic features resulting from CEF excitations and the Kondo effect in YbPtBi. Overall, this analysis places the `super'-heavy YbPtBi in the limit of $T_\text K>T_{\text{CEF},1}$ and provides strong support for the notion that the extremely high $\gamma$ value results from the hybridization of conduction electrons with a quasisextet CEF ground state.

The elastocaloric effect $\Delta T/\Delta\epsilon$ describes a temperature change $\Delta T$ that is induced by varying the strain by $\Delta\epsilon$. Thermodynamically it is given by 
 \begin{equation}
\frac{\Delta T}{\Delta\epsilon}
=
-\frac{\left.\partial S/\partial\epsilon\right|_T}{\left.\partial S/\partial T\right|_\epsilon}
=
-\frac T{C_V}\left.\frac{\partial S}{\partial\epsilon}\right|_T,
\label{eq:elastocaloric-definition}
\end{equation}
with $S$ being the entropy and $C_V$ the heat capacity at constant volume $V$.

As recently established in Ref.~\cite{Ikeda19}, $\Delta T/\Delta \epsilon$ can be determined with high precision in an a.c. version of the technique by applying an oscillation with amplitude $\Delta\epsilon$ in piezo-driven uniaxial pressure cells~\cite{Barber19,Hicks14} and measuring the resulting $\Delta T$ using a thermocouple (see Fig.~\ref{fig:overview}\,(b)). $\Delta\epsilon$ is determined through a capacitive displacement sensor (not shown). A constant, finite strain $\epsilon$ can be superimposed, so that $\Delta T/\Delta\epsilon$ can be mapped out as a function of $\epsilon$. In this work, we follow this novel experimental method to determine $\Delta T/\Delta\epsilon$ for YbPtBi with uniaxial pressure $p$ applied along the crystallographic $[1\,0\,0]$ direction, resulting in a finite $\epsilon$. We implemented following important modifications to the technique: first, we use a uniaxial pressure cell that also incorporates a force sensor~\cite{Barber19} since the applied force/pressure $p$ is a better control parameter than the conjugated strain $\epsilon$ in these type of devices. Second, we mount the sample free-standing in a sample carrier which is designed in such a way that only compression (denoted by a negative sign of $p$) can be applied when the mechanical contact in the carrier is closed (see Fig.~\ref{fig:overview}\,(b), Ref.~\cite{Jerzembeck22}). This allows us to determine precisely the neutral point $p=0$ (and $\epsilon=0$) at any given temperature which is important for the symmetry arguments presented below.

\begin{figure}
\centering
\includegraphics[width=\columnwidth]{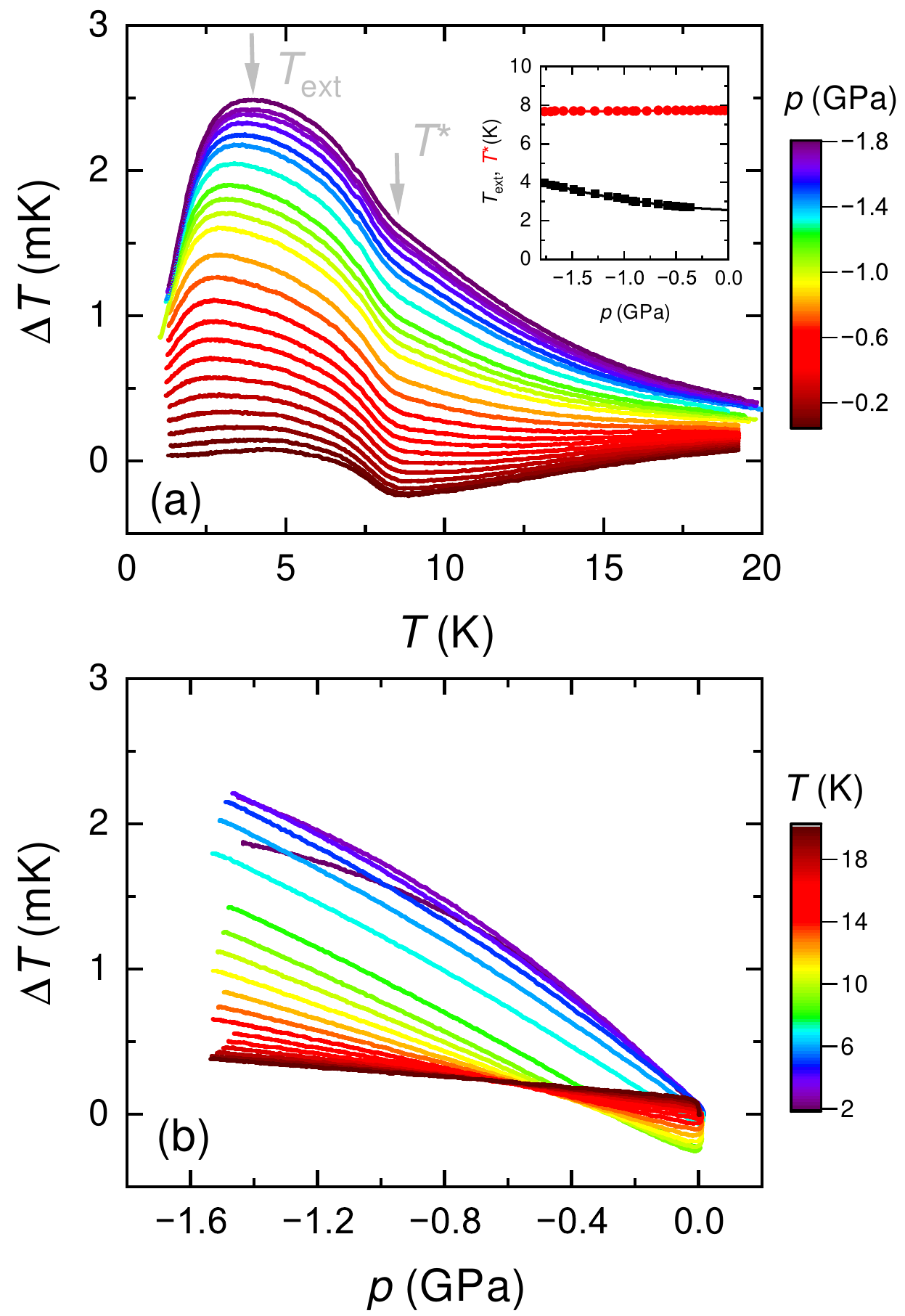} 
\caption{Elastocaloric temperature amplitude $\Delta T$ induced by a small oscillating uniaxial strain with amplitude $\Delta \epsilon\approx\rm const.<0$ at constant offset uniaxial pressures, $p$, as a function of temperature, $T$, (a) and as a function of $p$ at constant $T$ (b). Note that the $p\approx0$ data is also shown in Fig.~\ref{fig:overview} on a logarithmic $T$ scale. The data spacing in (a) is $\sim0.1\,\rm GPa$ and in (b) $1\,\rm K$ (for $T<10\,\rm K$) and $2\,\rm K$ (for $T>10\,\rm K$). The arrows in panel (a) indicate the position of the characteristic temperatures $T_\text{ext}$ and $T^\star$. The inset to (a) shows the evolution of $T_\text{ext}$ and $T^\star$ with $p$ (for criteria, see SI).}
\label{fig:elastocaloric}
\end{figure}


To illustrate the fingerprint of relevant energy scales in YbPtBi at ambient pressure in our data, we first compare on the right of Fig.~\ref{fig:overview} the elastocaloric temperature amplitude $\Delta T(T)$ for $T\gtrsim1.2\,\rm K$ (panel (e)) at ambient pressure $p_a\approx0$ and $\Delta\epsilon\approx\text{const.}$ with literature data~\cite{Mun13} on the molar specific heat $C_p(T)\propto T\left.\partial S/\partial T\right|_{p_a}$ (panel (c)) and the thermal expansion $\alpha(T)\propto\left.\partial S/\partial p\right|_T\propto\left.\partial S/\partial\epsilon\right|_T$ (panel (d)) (see SI for a discussion of the equivalence of different thermodynamic quantities at ambient pressure). Upon cooling, $\Delta T$ shows a clear feature around $T^*\approx7.6\,\rm K$  (see SI for criterion) with a concomitant sign change of $\Delta T$. $\alpha(T)$ also exhibits a similar feature as $\Delta T(T)$, including a sign change, at a slightly lower temperature. Simultaneously, the magnetic contribution $C_\text m$ to the specific heat $C_p$, obtained after subtracting the specific heat of the non-moment bearing Lu analogue\cite{Mun13}, shows a clear peak at $T^*$. Previously, these prominent features in $\alpha$ and $C_\text m$ were interpreted either to be solely due to CEF effects or to combined CEF/Kondo effects. Here, we will provide an alternative interpretation, and we will show that the first excited CEF level is in fact located much lower in energy. Upon further cooling down to $\approx1.2\,\rm K$, the lowest temperature of our experiment, $\Delta T(T)$ remains featureless and small, similar to $\alpha(T)$. At even lower temperatures $C_p(T)$ and $\alpha(T)$ show clear features associated with a magnetic phase transition at $T_\text N\approx0.4\,\rm K$.

Now we turn to the behavior of $\Delta T$ under finite, symmetry-breaking uniaxial pressure $p$ up to $\approx-1.8\,\rm GPa$ compression, shown in Fig.~\ref{fig:elastocaloric}. Clearly the data sets as a function of $T$ (Fig.~\ref{fig:elastocaloric}\,(a)) and $p$ (Fig.~\ref{fig:elastocaloric}\,(b)) reveal significant changes of $|\Delta T|$. Key observations can be summarized as follows: first, as $|p|$ is increased, a low-temperature extremum emerges at $T_\textrm{ext}\approx2.7\,\rm K$ for $p\approx-0.3\,\rm GPa$, which is increased to $4.0\,\rm K$ by $p\approx-1.8\,\rm GPa$. Second, the feature at $T^*\approx7.6\,\rm K$ remains visible for all pressures and its position is only barely affected by $p$ (see inset of Fig.~\ref{fig:elastocaloric}\,(a)). Third, $\Delta T$ increases strongly, from almost zero, in a monotonic and, to first approximation, in a near linear manner with $|p|$ for any given $T$. Only for larger $|p|\gtrsim0.5\,\rm GPa$ are some deviations from this $p$-linear behavior, in particular at lowest $T$, observed.

\begin{figure*}
\centering
\includegraphics[width=17.8cm]{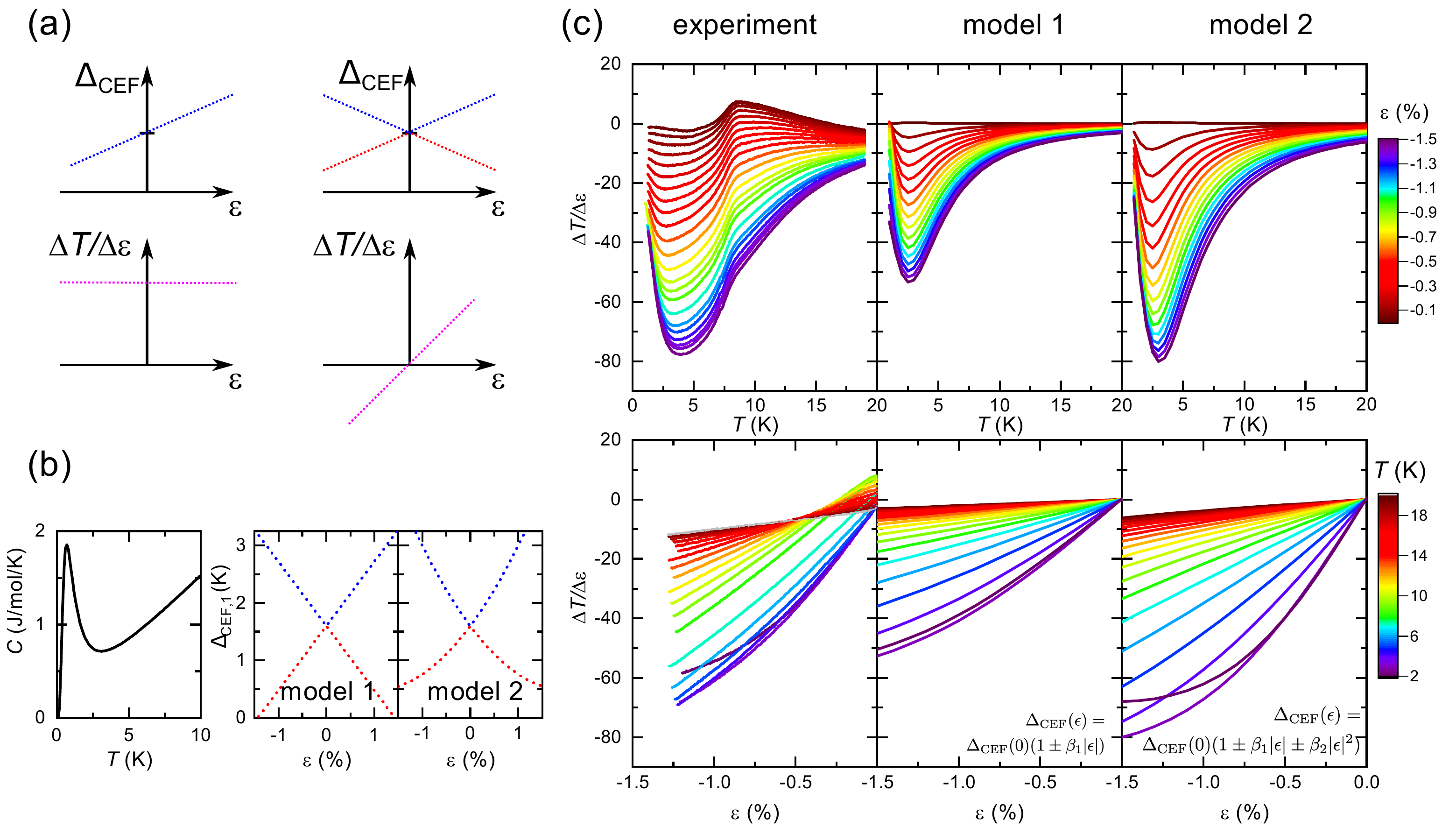} 
\caption{(a) Modelling of the impact of a strain $\epsilon$ on the CEF energy level difference $\Delta_\text{CEF}$ and the resulting behavior of $\Delta T/\Delta\epsilon$ at constant temperature $T={\cal O}(\Delta_\text{CEF})/k_\text B$. Left: scenario (i), induced shift of the first excited level. Right: scenario (ii), induced splitting of the first excited level. (b) Left: model specific heat $C_V(T)$ at $\epsilon=0$. Right: splitting of $\Delta_\text{CEF}$ as a function of $\epsilon$ used in model 1 and model 2, both within scenario (ii) (see text for details). (c) Comparison of experimental data of the elastocaloric effect and the results of the two model calculations. For the experimental data, $Y_{100}\approx120\,\rm GPa$ was used~\cite{Oguchi01} to express $p$ in terms of $\epsilon$ (see text and SI). The data spacing for both experimental data and model calculations in the top panel is $\sim\,0.08$\% and in the bottom panel $1\,\rm K$.}
\label{fig:modelling}
\end{figure*}

In general, this elastocaloric effect data contains contributions from all relevant energy scales, in particular the CEF and Kondo energy scales. In the following we will use generic qualitative arguments and explicit modelling of the elastocaloric effect of single-ion CEF states to disentangle these contributions. In particular, we will demonstrate that the strong change of temperature $\Delta T$ with uniaxial pressure $p$ results from the response of the first excited quartet CEF level to symmetry breaking, whereas the behavior of $\Delta T$ at $p=0$ (including the anomaly at $T^\star$) most likely originates from the formation of the coherent Kondo state.

To facilitate the discussion of the elastocaloric effect of CEF levels, we will from now on focus on the notion of strain $\epsilon$, since Young's modulus $Y_{100}:=\partial p/\partial\epsilon|_T\approx\text{const.}$ (see SI), and assume temperatures $T={\cal O}(\Delta_\text{CEF})/k_\text B$ where $\Delta_\text{CEF}$ is the energy difference between ground state and first excited CEF energy level. The two dominant effects of finite $\epsilon$ are expected to be (i) shifting the CEF levels and (ii) a possible lifting of degenerate CEF levels due to lowering the crystal symmetry. The latter scenario can only occur when the degenerate state is not the ground state, since the degeneracy would otherwise be lifted at zero strain through a spontaneous Jahn-teller distortion. The elastocaloric effect $\Delta T/\Delta\epsilon$ is expected to be significantly distinct in these two cases.

The two scenarios are visualized separately in Fig.~\ref{fig:modelling}\,(a): In case (i), left sketch of the figure, the CEF energy level is uniquely associated with $\epsilon$, necessary for the applicability of Grüneisen scaling $\partial p/\partial T|_V\propto C_V/V$ (see SI). We obtain $\partial S/\partial\epsilon|_T\propto T\partial S/\partial T|_\epsilon$, therefore $\Delta T/\Delta\epsilon\approx\text{const.}$ as a function of strain, see Eq.~(\ref{eq:elastocaloric-definition}). Thus, in this case, we expect a large intercept of $\Delta T/\Delta\epsilon$ at $\epsilon=0$ and no significant change with $\epsilon$.

In case (ii), the strain-induced symmetry lowering leads to a splitting of the first excited CEF energy level for both compressive and tensile strains. Hence, at $\epsilon=0$ we must have $\partial S/\partial\epsilon|_T=0$, correspondingly $\Delta T/\Delta\epsilon=0$ and Grüneisen scaling is no longer applicable. We also must have $\partial S/\partial\epsilon|_T\propto\epsilon$. Therefore, in scenario (ii), we expect that the magnitude of $\Delta T/\Delta\epsilon$ is expected to increase rapidly from a small value at $\epsilon=0$ with increasing tension or compression (see right sketch of Fig.~\ref{fig:modelling}\,(a)).

Our data (see Fig.~\ref{fig:elastocaloric} and \ref{fig:modelling}(c)) is characterized by a large change of $\Delta T/\Delta \epsilon$ with $\epsilon$ and only a small finite intercept at $\epsilon=0$. We can therefore conclude that the elastocaloric effect under finite strains is dominated by a strain-induced splitting of a first excited CEF level. This CEF level has to be the $\Gamma_8$ quartet, since the Yb$^{3+}$ Kramers doublets are protected by time-reversal symmetry and cannot be split by the application of strain.

We note that these considerations also imply that measurements of the thermal expansion $\alpha$ at ambient pressure will only display anomalies of excited CEF levels when these levels are doublets that only shift with strain. In contrast, the excited CEF $\Gamma_8$ level will leave almost no fingerprint in $\alpha(p=0)\propto\left.\partial S/\partial\epsilon\right|_T(\epsilon=0)\approx0$ when the level splitting is the dominant effect. Therefore, a finite $\alpha$ seems unlikely to be related to the physics of the $\Gamma_8$ level, contrary to what has been discussed in previous studies on YbPtBi~\cite{Ueland15} under the assumption of validity of Grüneisen scaling (see SI).

To extract quantitative information on the CEF states from the elastocaloric data, in particular on the estimate of energy of the $\Gamma_8$ level in YbPtBi, we performed model calculations of $\Delta T/\Delta\epsilon$ using a Schottky-type specific heat $C_p$ of a two-level system with four-fold degenerate first excited state at energy $\Delta_\text{CEF}$.  In addition, since our measured signal is $\Delta T\propto1/C_p$, we added an electronic contribution of the form $\gamma T$ to the total model specific heat $C_p$. To reduce the number of parameters, we omitted phononic contributions because they are small below $\sim10\,\rm K$ (see Fig.~\ref{fig:overview}(c)). The full model specific heat at $\epsilon=0$ is shown in the left panel of Fig.~\ref{fig:modelling}\,(b). Using a value of $\Delta_\text{CEF}/k_\text B=1.6\,\rm K$, this model reproduces the broad hump in the literature $C_P(T)$ data~\cite{Fisk91,Mun13} around $T\approx800\,\rm mK$.

To parameterize the response to strain, we use two approximations, which we call model~1 and model~2. Model~1 comprises a linear splitting by strain via $\Delta_\text{CEF}(\epsilon)=\Delta_\text{CEF}(0)\left(1\pm\beta_1|\epsilon|\right)$. The choice for this model is motivated by considering the effect of a tetragonal distortion on the CEF-Hamiltonian~\cite{Lea62} ${\cal H}_\text{CEF}^\text{cubic}\to{\cal H}_\text{CEF}^\text{cubic}+g_{zz}\epsilon O_2^0$ perturbatively with an elastic constant $g_{zz}$ that characterizes the distortion and $O_2^0$ the Stevens operator that emerges in tetragonal symmetry (see SI). The energy of the $\Gamma_8$ state then changes as $E_{\Gamma_8}\rightarrow E_{\Gamma_8}\pm6g_{zz}\epsilon$ for small $\epsilon$.

Naturally deviations from the $\epsilon$-linear behavior of the CEF excitation energy will arise for larger $|\epsilon|$. In an attempt to better describe the magnitude of our experimental data,  we include a second-order term of the expansion in $\epsilon$ in model~2. The corresponding energy splitting we use then reads $\Delta_\text{CEF}(\epsilon)=\Delta_\text{CEF}(0)\left(1\pm\beta_1|\epsilon|\pm\beta_2|\epsilon|^2\right)$.

Figure~\ref{fig:modelling}\,(c) shows a comparison of the experimental data for $\Delta T/\Delta\epsilon$ (left column) to the calculations for model 1 (middle column) and model 2 (right column). For both models we use $\Delta_\text{CEF}(0)/k_\text B=1.6\,\rm K$ and $\beta_1=70$. Since the lowest temperature of our experiments is $T_\text{min}\approx1.2\,\rm K$, we restrict the calculations to this temperature range. Therefore, the model results reflect the effects associated with the higher-energy branch (blue dotted lines in the right panel of Fig.~\ref{fig:modelling}\,(b)).

Clearly, the results for model~1 already capture many of the essential observations of the experiment on a qualitative level. It reproduces the low-temperature minimum of $\Delta T/\Delta\epsilon$ as a function of $T$ around $T\approx3\,\rm K$ under finite $\epsilon$, as well as the approximately linear change of $\Delta T/\Delta \epsilon$ with $\epsilon$.

However, model~1 is not sufficient to account for the data also on a quantitative level. Taking into account a second-order term with model~2 and repeating our calculations with $\beta_2=1750$, we obtain the results shown in the right column of Fig.~\ref{fig:modelling}\,(c). Now much of the experimental data can be very well reproduced over the full strain range.

It is important to note that the good agreement between experiment and model calculations is only achieved when $\Gamma_8$ is the first excited state (scenario (ii)). In the SI, we also show model calculations for the reverse scenario (we call it scenario (iii)) assuming $\Gamma_8$ is the ground state and the first excited state is a doublet, even if this scenario is unlikely due to the inherent instability of a symmetry-protected degeneracy of the ground state towards Jahn-Teller distortions. We find that scenario (ii) and (iii) can be clearly distinguished based on our experimental data by considering the value of $T_\text{ext}$ for $\epsilon\rightarrow0$. Specifically, we find that the finite value of $T_\text{ext}$ for $\epsilon\rightarrow0$, shown in the inset of Fig.~\ref{fig:elastocaloric}(a), is only compatible with scenario (ii).

 Overall, our analysis establishes the $\Gamma_8$ quartet to be the first excited state with an energy difference to the doublet ground state of $\Delta_\text{CEF}\approx k_\text B\cdot1.6\,\rm K$. However, the physics of single-ion CEF levels does not capture the feature at $T^\star\sim 7.5\,\rm K$ in $\Delta T/\Delta\epsilon$ which persists for all uniaxial pressures (see Fig.~\ref{fig:elastocaloric}\,(a) and Fig.~\ref{fig:modelling}\,(c)). The model would also predict that $\Delta T/\Delta\epsilon\to0$ for $\epsilon\to0$, required by restoring the cubic crystal symmetry. Instead we observe a small but finite $\Delta T/\Delta\epsilon$, which we attribute to contributions from other energy scales than the CEF $\Gamma_8$ one.
 
 Since the remaining doublet CEF level excitation is located at much higher temperatures around $60\ldots100\,\rm K$, it can most likely be excluded as the source for the anomaly at $T^\star$. The obvious temperature scale which is known to be relevant in YbPtBi is set by the Kondo temperature $T_\text K$. Remarkably, the thermal expansion data~\cite{Mun13} shown in Fig.~\ref{fig:overview}\,(d) reveals the onset of negative $\alpha$ below $T\approx20\,\rm K$ that persists down to $T\approx T^\star$. In general, a negative thermal expansion in cubic systems is exceptional~\cite{Mazzone20}. To rationalize this, we note that the volume of Yb$^{3+}$ is smaller than the one of Yb$^{2+}$. Therefore, even tiny hybridization-induced changes~\cite{Brandt84,Kummer18} of the strictly trivalent state of Yb can be the origin of a negative $\alpha$. Thus, all experimental data are consistent with the expectations for the formation of the Kondo lattice with $T_K\approx T^\star\approx10\,\rm K$.

The hierarchy of temperature scales in YbPtBi can now be clearly assigned to $T_{\text{CEF},2}>T_\text K>T_{\text{CEF},1}$. Therefore, this opens the possibility that the conduction electrons do not only hybridize with the Yb$^{3+}$ CEF $\Gamma_7$ doublet ground state, but also with the first excited $\Gamma_8$ quartet state. Effectively, hybridization then takes place with an Yb quasi-sextet ($N=6$). In fact, the analysis~\cite{Tsujii05} of the generalized Kadowaki-Woods ratio $A/\gamma^2$, with $A$ being the Fermi-liquid coefficient of the resistivity, shows that $\rho(T)$ for YbPtBi~\cite{Fisk91,Torikachvili07} falls between the values of $A$ expected for $N=6$ and $N=8$. This is nonetheless remarkable since the hybridization strength can also depend on the symmetry of the underlying CEF level~\cite{Vyalikh10,Dong14,Rahn22}, an aspect which has so far been rarely considered. Our study shows that YbPtBi with two CEF levels of different symmetry below $T_\text K$ might be an interesting reference system to quantify the relevance of symmetry-dependent hybridization strength.


In summary, through measurements and analyses of the elastocaloric effect, we have firmly established the hierarchy of energy scales in the `super'-heavy fermion material YbPtBi. We find that the Kondo energy $k_\text BT_\text K\approx k_\text B\cdot10\,\rm K$ is higher than the energy difference $\Delta_\text{CEF}$ between the ground state and the first excited quartet CEF level with $\Delta_\text{CEF}\approx k_\text B\cdot1.5\,\rm K$, putting both the $\Gamma_7$ ground state doublet and the $\Gamma_8$ quartet below $k_\text BT_\text K$.  This allows for the possibility that conduction electrons hybridize with a quasi-sextet ($N=6$) Yb$^{3+}$ ground state, providing strong support for theoretical models that assign the anomalously large electron mass to an enhanced degeneracy of the CEF levels.

At a more general level, our work demonstrates that measurements of the elastocaloric effect under finite pressures~\cite{Ikeda21,Li22,Ye22,Rosenberg19}  enable us to control and quantify strain-induced changes of the crystal-electric field schemes and disentangle relevant low-energy scales in correlated electron systems in a novel way. This approach will also be particularly relevant for the field of quantum magnets, in which the unambiguous determination of single-ion CEF states is essential for a microscopic description of their unusual magnetic properties.

\textit{Methods - }Single crystals of YbPtBi were grown from a Bi-riched ternary melt following the procedure described in Ref. \cite{Canfield91,Mun13,Canfield20} and in the SI. The samples were polished for measurements under finite uniaxial pressures\cite{Barber19} into a bar with dimensions of $100\,\mu{\rm m}\times140\,\mu{\rm m}\times1000\,\mu{\rm m}$, with the long axis being the strain axis. For measurements of the a.c. elastocaloric measurements, the d.c. voltages on the piezoelectric actuators were modulated by a small a.c. voltage on the tension stack. For the measurements of the induced temperature change $\Delta T$, an chromel–AuFe$_{0.07\%}$ thermocouple \cite{Stockert11} was fixed to the sample with a tiny amount of Stycast 1266. The thermocouple was anchored on the cell body. The voltage on the thermocouple was amplified by a low-temperature transformer mounted on the low-temperature stage and subsequently read out by a Lock-In amplifier. Further description on the uniaxial pressure cell and modelling of the elastocaloric effect are included both in the main text (see Fig.~\ref{fig:overview}) and the SI.

\textit{Acknowledgments -} We acknowledge useful discussions with P. Thalmeier and thank E. Mun and B. Kuthanazhi for providing the ambient-pressure thermodynamic data. We also acknowledge the Gordon and Betty Moore foundation for funding the International Workshop "Experimental Advances in the Use of Pressure and Strain to Probe and Control Quantum Matter", which initiated the idea for this project.  PCC acknowledges G. Wells for not letting YbPtBi be called a "morbidly obese Fermion". Financial support by the Max Planck Society is gratefully acknowledged. In addition, we gratefully acknowledge funding through the Deutsche Forschungsgemeinschaft (DFG, German Research Foundation) through TRR 288—422213477 and the SFB 1143 (project-id 247310070). Research in Dresden benefits from the environment provided by the DFG Cluster of Excellence ct.qmat (EXC 2147, project ID 390858940). Work at the Ames National Laboratory was supported by the U.S. Department of Energy, Office of Science, Basic Energy Sciences, Materials Sciences and Engineering Division. The Ames National Laboratory is operated for the U.S. Department of Energy by Iowa State University under Contract No. DEAC02-07CH11358.

\bibliographystyle{apsrev}

\clearpage

\section{Supplemental Information}

\subsection{Crystal growth details}

Single crystals of YbPtBi were grown from ternary melts rich in Bi \cite{Canfield20}.  High purity, elemental Yb, Pt and Bi were combined in an alumina crucible in a 10(Yb):10(Pt):80(Bi) atomic ratio, sealed in an amorphous silica tube under 1/6 atmosphere of Ar and heated to 1100 $^\circ$C over 4 hours.  After dwelling at 1100 $^\circ$C for 3 hours, the ampule was cooled to 900 $^\circ$C over 2 hours and then slowly cooled to 600 $^\circ$C over 231 hours.  After dwelling at 600 $^\circ$C for a few hours the ampule was taken out of the furnace, inverted into a metal centrifuge cup and rotated a several thousand rpm for times less than 15 seconds  \cite{Canfield20} so as to separate the remaining, very Bi rich, remaining liquid from the grown YbPtBi crystals.  The resultant crystals were well faceted and could be as large as 5-10 mm on a side.  Given that YbPtBi is air sensitive, crystals were sealed in evacuated silica tubes for long term storage.

\subsection{Determination of characteristic temperatures}
\label{sec:criteria}

\begin{figure}[h]
\centering
\includegraphics[width=\columnwidth]{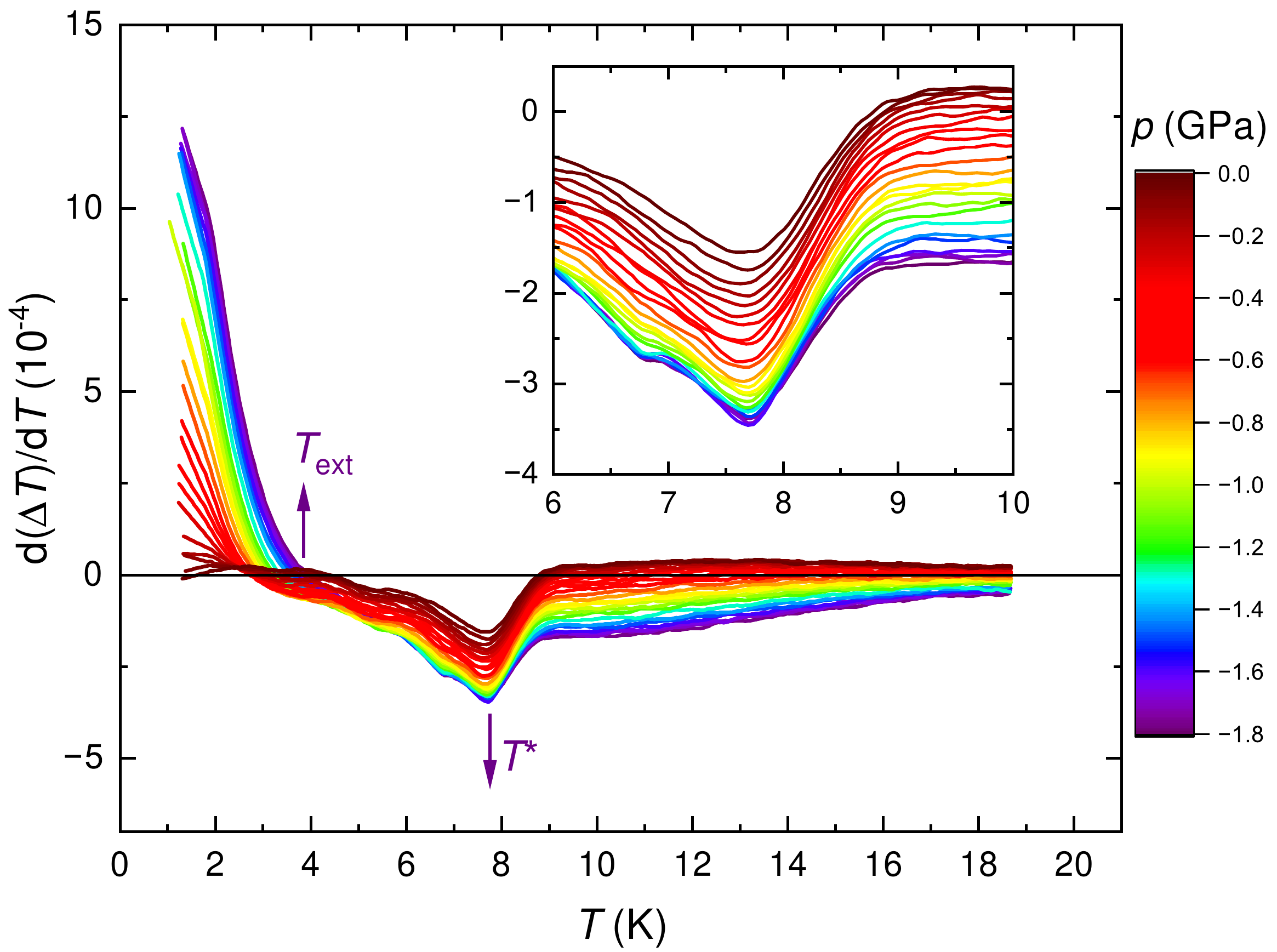} 
\caption{Determination of characteristic temperatures from our elastocaloric effect data, $\Delta T$. Main panel: Derivative of $\Delta T$ with respect to temperature, $\left.\partial\Delta T/\partial T\right|_p$, at constant pressures. Characteristic temperatures $T_\text{ext}$ and $T^\star$ are indicated by arrows for the data set at highest compression $p=-1.78\,\rm GPa$. Inset: Data on enlarged scales.}
\label{fig:criteria}
\end{figure}

In the main text, we introduce two characteristic temperatures $T_\text{ext}$ and $T^\star$ in our elastocaloric effect data $\Delta T$ as a function of $T$ (see Fig.~2 of the main text). To clearly define the criteria to infer $T_\text{ext}$ and $T^\star$, we show in Fig.~\ref{fig:criteria} the derivative of $\Delta T$ with respect to temperature at constant uniaxial pressures. Since $T_\text{ext}$ marks the position of the extremum in $\Delta T$, which emerges for higher compression, we determine it from the low-temperature zero-crossing of d$(\Delta T)$/d$T$. The characteristic temperature $T^\star$ is associated with a feature in $\Delta T$, that gives rise to a clear minimum in d$(\Delta T)$/d$T$. We use the position of the minimum at each $p$ to determine $T^\star(p)$. The evolution of $T_\text{ext}(p)$ and $T^\star(p)$ are shown in the inset of Fig.~2 in the main text. Whereas $T^\star$ shows only a weak $p$ dependence, $T_\text{ext}$ is best described by a quadratic function with a zero-strain intercept of $\approx2.6\,\rm K$.

\subsection{Direct comparison of thermodynamic quantities}

\begin{figure}[h]
\centering
\includegraphics[width=\columnwidth]{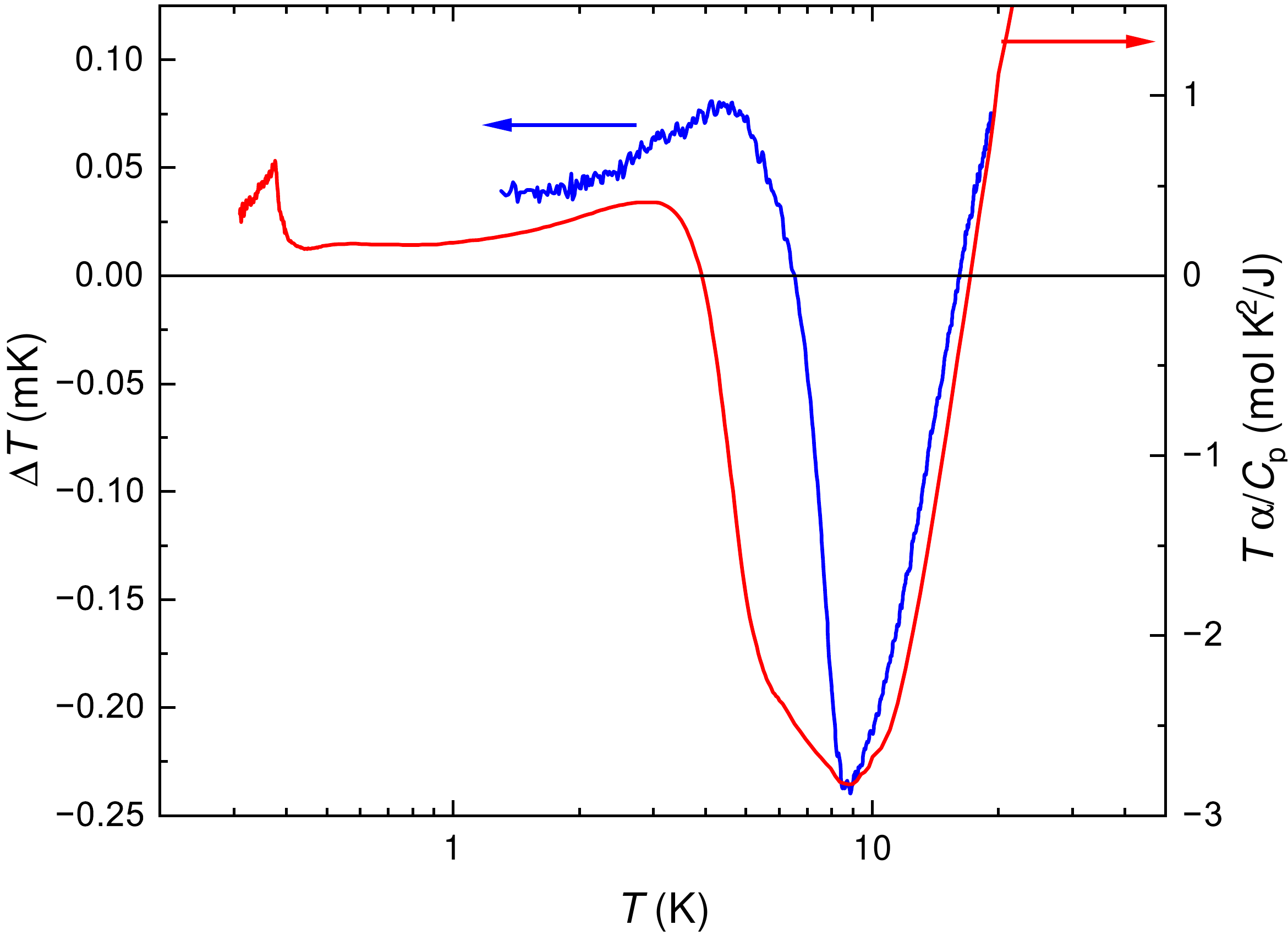} 
\caption{Direct comparison of elastocaloric effect amplitude, $\Delta T$ (left panel), to the combination of thermodynamic quantities $T\alpha/C_p$ (right panel), which is calculated from literature data \cite{Mun13}.}
\label{fig:DT-comparison}
\end{figure}

The various thermodynamic quantities, discussed in the main text, are defined as
\begin{align}
   \frac{\Delta T}{\Delta\epsilon}
    &=
    -\frac T{C_V}\left.\frac{\partial S}{\partial\epsilon}\right|_T,
    \nonumber\\
    C_p
    &=
    T\left.\partial S/\partial T\right|_{p_a},
    \nonumber\\
    \alpha(T)
    &\propto
    \left.\partial S/\partial p\right|_T
    \propto
    \left.\partial S/\partial\epsilon\right|_T
\end{align}
with $\Delta T/\Delta\epsilon$ the elastocaloric effect, $C_p$ the heat capacity and $\alpha$ the thermal expansion coefficient.\footnote{The thermal expansion coefficient here is defined using the notion that compression is denoted by a negative pressure.} These equations suggest that, in first approximation, $\Delta T/\Delta\epsilon\propto-T\alpha/C_V\approx-T\alpha/C_p$. We therefore compare in Fig.~\ref{fig:DT-comparison} the $\Delta T$ data, measured in this work at $p\approx0$ and with an oscillation amplitude $\Delta \epsilon<0$, with literature data on $T\alpha/C_p$, calculated from data presented in Ref.~\cite{Mun13}, explicitly. We see that both quantities display similar features and also the proportionality is, in first approximation, obeyed. The only marked difference is a difference in the characteristic temperature, $T^\star$. The origin of this temperature difference is not clear at present. However, it is important to note that this temperature difference does not have any impact for the conclusions drawn in the main text, since we only discuss the evolution of $T^\star$ with finite, uniaxial $p$.

\subsection{Calibration of the elastocaloric signal}

For elastocaloric effect measurements, an independent calibration of $\Delta T/\Delta \epsilon$ is often needed, since finite heat flow out of the sample in the pressure-cell environment require correction factors for the magnitude of $\Delta T$ (see Ref.~\cite{Straquadine20} for finite-element simulations of the heat flow). In our case, we performed this calibration in the following way. The pressure dependence of the characteristic temperature $T^\star$ is determined in our work with high precision to be $\Delta T^\star/\Delta p\approx(0.03\pm0.01)\,\rm K/GPa$ (see Fig.~\ref{fig:criteria} and Fig.\,2 in the main text). From thermodynamic considerations, this slope has to be identical to the absolute value of $1/Y_{100}\left( \Delta T/\Delta\epsilon|_{T\to T^\star_+}-\Delta T/\Delta \epsilon|_{T\to T^\star_-}\right)$ (Ehrenfest relation). This procedure yields a scaling factor of $\approx1.5$, which has to be applied to the experimentally measured $\Delta T/\Delta \epsilon$. This factor is similar to the one used in previous works \cite{Li22,Ikeda19}. All data shown in the main text are rescaled by this scaling factor.

\subsection{Young's modulus of YbPtBi}

\begin{figure}[h]
\centering
\includegraphics[width=\columnwidth]{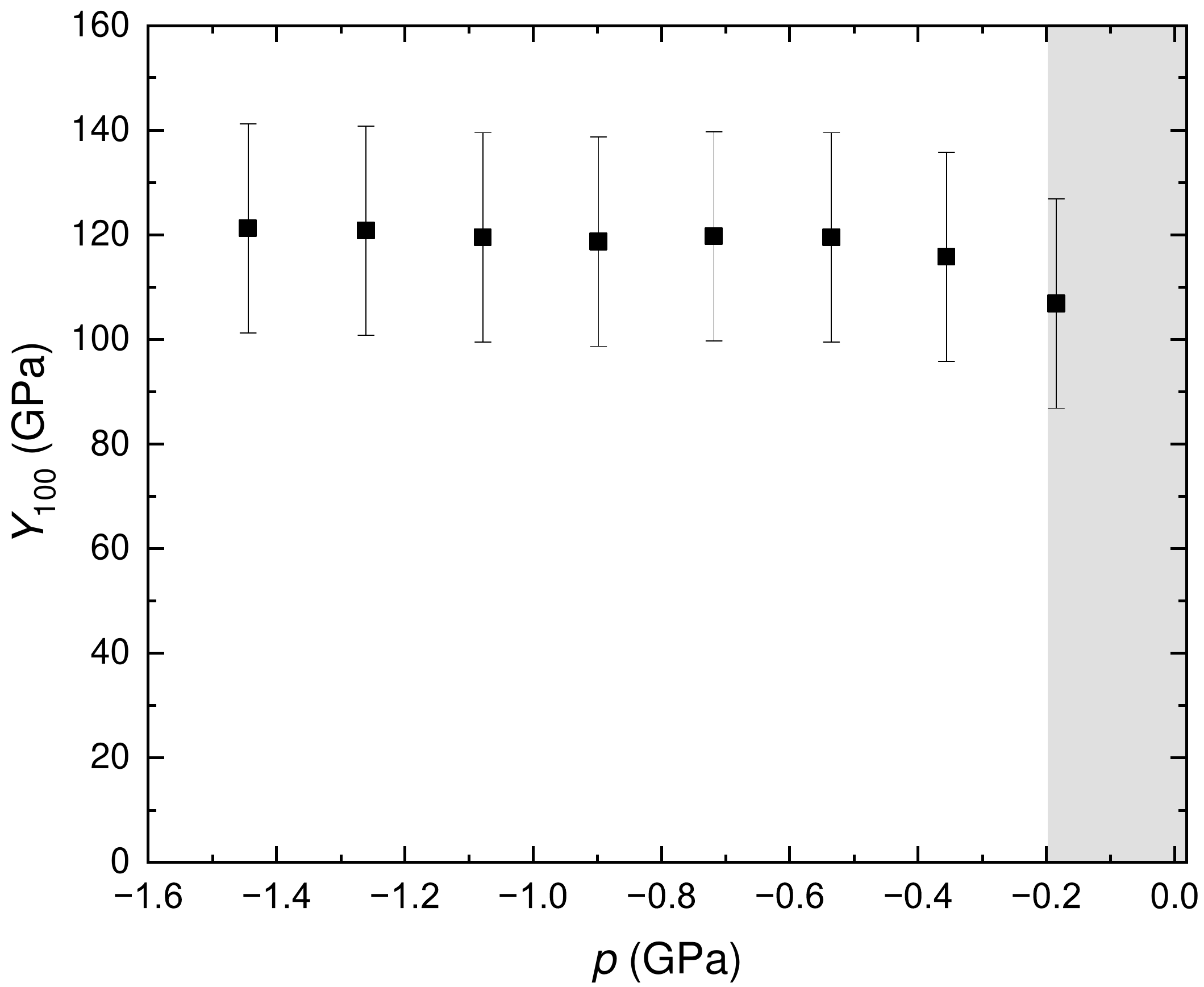} 
\caption{Estimate of the Young's modulus of YbPtBi along the $(1\,0\,0)$ direction, $Y_{100}$ as a function of pressure $p$ at a temperature of $4\,\rm K$. Data in the grey-shaded region is dominated by closing of the mechanical contact in the sample carrier, and therefore has been omitted. The error bars reflect the systematic error in the determination of geometric factors and other extrinsic contributions to the measurements of the displacement and force sensors.}
\label{fig:Youngmodulus}
\end{figure}

In the main text, we use the assumption that $Y_{100}:=\partial p/\partial\epsilon|_T\approx\text{const.}$ In addition, we use a calculated literature value of $120\,\rm GPa$, estimated for LaPtBi \cite{Oguchi01}. Given that we used a uniaxial-pressure cell with integrated force and displacement sensors \cite{Barber19}, it is in principle possible to determine the Young's modulus of YbPtBi in our experiment simultaneous to the determination of its elastocaloric effect. An example data set of the estimated $Y_{100}$ vs. $p$, taken at a temperature of $T=4\,\rm K$, is shown in Fig.~\ref{fig:Youngmodulus}. This data set shows that $Y_{100}\approx120\,\rm GPa$ and constant as a function of $p$ and $\epsilon$. In addition, we did not observe a significant temperature dependence of $Y_{100}$ below $20\,\rm K$. Thus, the assumptions of the main text are justified.

\subsection{Validity of Gr\"uneisen scaling and implications for interpretations of ambient-pressure thermal expansion measurements of excited CEF levels}

In the main text, we discuss different scenarios for the behavior of CEF levels under strain and argue that Gr\"uneisen scaling does not apply in all cases, in particular when the excited CEF level is allowed to split with strain by symmetry. We argue that this has the important consequence that thermal expansion measurements at ambient pressure will not show any signatures of such a CEF level. Here, we want to expand on these arguments, giving more details.

The basis of Gr\"uneisen scaling in any given system is that its physics is dominated by a single temperature scale $T^\star=T^\star(\epsilon)$ (for example the Debye temperature for phonons or the Fermi temperature for metals). With $x:=T/T^*$ we then may write $S(T,\epsilon)\to S(x)$ and obtain
\begin{equation}
    \left.\frac{\partial S}{\partial\epsilon}\right|_T
    =
    \left.\frac{\partial S}{\partial T}\right|_\epsilon
    \cdot
    \frac{\partial x/\partial\epsilon|_T}{\partial x/\partial T|_\epsilon}
    =
    -T\left.\frac{\partial S}{\partial T}\right|_\epsilon
    \frac1{T^*}\left.\frac{\partial T^*}{\partial\epsilon}\right|_T.
\end{equation}
Using definitions for $\alpha$ and $C_V$ given in the main text, this leads to $\alpha\propto\Gamma C_V$, with $\Gamma=(1/T^*)\left.\partial T^*/\partial\epsilon\right|_T$ the famous Gr\"uneisen parameter (see e.g. \cite{Zhu03}). Since $\Gamma$ is to a good approximation independent of temperature, this lead to the famous Gr\"uneisen scaling relation $\alpha\propto C_V$.

After these general thermodynamic remarks, we return to the discussion of the implications for the analysis of excited CEF levels. Similar to the discussion presented in the main text, we will distinguish between the case in which a single excited CEF level only shifts in energy with $\epsilon$ (scenario (i)), and the case in which an excited CEF level is allowed to split into two branches with $\epsilon$ (scenario (ii)). In scenario (i), $T^\star$ can be clearly assigned to a single value of $\epsilon$, and thus, Gr\"uneisen scaling is expected to be valid, since $\Gamma$ is well-defined, finite and temperature-independent. This implies that the thermal expansion at ambient pressure, $\alpha(T)$, will show a Schottky-type anomaly related to the CEF energy, similar to $C_V(T)$. In contrast, in scenario (ii) the Gr\"uneisen parameter is ill-defined since $T^\star$ is not clearly assigned to a single value of $\epsilon$. The most important consequence is that Gr\"uneisen scaling does not apply in this scenario, and thus that $\alpha(T)$ and $C_V(T)$ at ambient pressure will be distinctly different. In particular, due to the symmetric change of $T^\star$ upon compression ($\epsilon<0$) and tension ($\epsilon>0$), $\partial S/\partial\epsilon|_T=0$ at $\epsilon=0$. Since $\alpha (p=0)\propto\partial S/\partial\epsilon|_T(\epsilon=0)=0$, the contribution of an excited CEF level, that splits with strain, to $\alpha$ has to be zero at ambient pressure. Thus thermal expansion at ambient pressure is not a suitable tool to detect the excited CEF levels in scenario (ii). Only under finite strains, such as is the case in the present study, $\partial S/\partial\epsilon|_T$ will not be zero.

\subsection{Model results for $\Gamma_8$ as first excited state vs. $\Gamma_8$ as ground state}

\begin{figure}[h]
\centering
\includegraphics[width=\columnwidth]{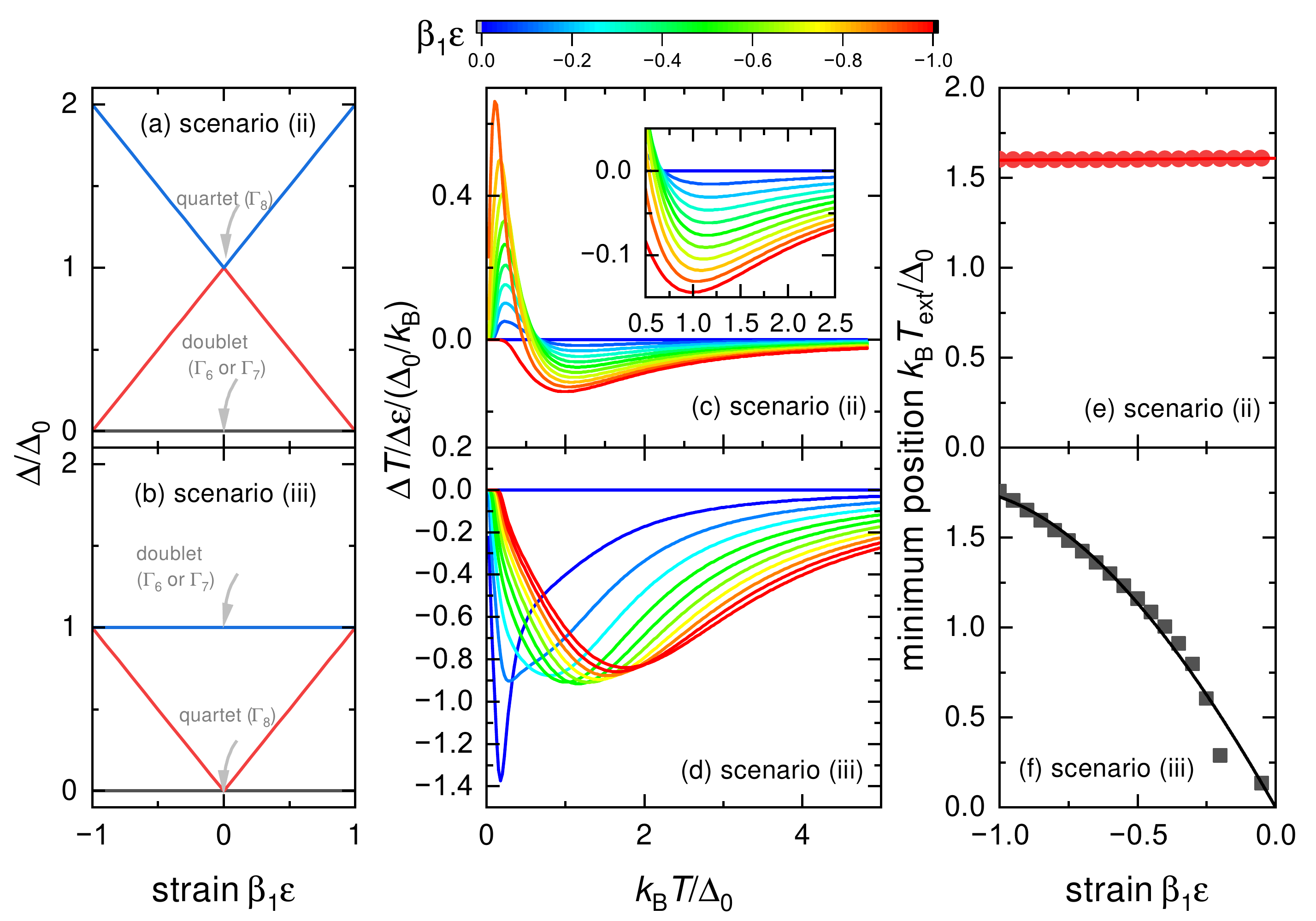} 
\caption{Contrasting the scenarios of quartet CEF ground state and doublet first excited CEF state (scenario (ii)) vs.  CEF ground state and doublet first excited CEF state (scenario (iii)). Schematic representations of the energy schemes in scenario (ii) (a) and scenario (iii) (b); Model calculations for the elastocaloric effect, $\Delta T/\Delta \epsilon/(\Delta_0/k_B)$, as a function of reduced temperature, $k_B T/\Delta_0$, at different strains, $\epsilon/\beta_1$, for scenario (ii) (c) and scenario (iii) (d); Position of the minima in the elastocaloric effect data as a function of strain for scenario (ii) (e) and scenario (iii) (f). Lines in (e) and (f) are guide to the eyes.}
\label{fig:ground-vs-first}
\end{figure}

In the main text, we discuss model calculations for the scenario (ii), in which the CEF ground state is a doublet ($\Gamma_6$ or $\Gamma_7$) and the first excited level is the quartet $\Gamma_8$ level (with an energy difference of $\Delta_0$ at zero strain). The $\Gamma_8$ quartet is allowed to split by strain, which to lowest order results in $\Delta(\epsilon)=\Delta_0\left(1\pm\beta_1|\epsilon|\right)$. For completeness, we contrast the model calculations of the elastocaloric effect within this scenario (ii) here with same calculations within a scenario (iii), in which $\Gamma_8$ is the ground state and the doublet the first excited state with energy gap $\Delta_0$ in zero strain. The ground state will then also split with strain, such that the lower branch remains the ground state at energy $\Delta(\epsilon)=0$ and the upper branch follows $\Delta(\epsilon)=\Delta_0\left(\beta_1|\epsilon|\right)$, and the doublet remains unchanged at $\Delta(\epsilon)=\Delta_0$. These scenarios are visualized in Figs.~\ref{fig:ground-vs-first} (a) and (b).

Figs.~\ref{fig:ground-vs-first} (c) and (d) show the model calculations of the elastocaloric effect for scenario (ii) and scenario (iii). Similar to the main text, we show the temperature dependence of the elastocaloric effect $\Delta T/\Delta \epsilon$ (in units of $\Delta_0/k_\textrm{B}$) at different compressive strains with $\epsilon<0$. Even though in both cases the entropy landscape is symmetric around $\epsilon=0$ and it is therefore expected that $\Delta T/\Delta\epsilon\propto\left.\partial S/\partial\epsilon\right|_T$ changes significantly with $\epsilon$ at a given $T$ (see main text), there are marked differences in the quantitative results of these model calculations. Within scenario (ii), for any finite strain, $\Delta T/\Delta \epsilon$ is positive for small $T$ and then changes sign at $k_\textrm{B} T\approx \Delta_0$. In contrast, for any finite strain in scenario (iii), $\Delta T/\Delta \epsilon$ is negative across the entire $T$ range up to at least $k_\textrm{B} T\approx 5 \Delta_0$.

In our experimental data, shown in Fig.~3 of the main text, $\Delta T/\Delta \epsilon$ is negative down to the lowest temperature measured, which is $1.2\,\rm K$, and the temperature dependence shows a clear minimum. The calculations for scenario (ii) indicate that such a minimum occurs for temperatures slightly higher than $\Delta_0/k_\textrm{B}$. In fact, in this scenario, the position of this minimum is to a good approximation unchanged with strain, as shown in Fig.~\ref{fig:ground-vs-first}\,(e). In contrast, the position of the minimum in $\Delta T/\Delta \epsilon(T)$ in the results for scenario (iii) is strongly strain-dependent, see Fig.~\ref{fig:ground-vs-first}\,(f). Particularly, the position of the minimum extrapolates to zero for $\epsilon\rightarrow 0$. As we show in the inset of Fig.~2 of the main text, the position of the minimum in our data extrapolates to a finite temperature of $\approx2.6\,\rm K$ and not to zero. We therefore conclude that our elastocaloric effect data is only compatible with scenario (ii), in which the $\Gamma_8$ quartet is the first excited state.

\subsection{Crystal-electric field Hamiltonian}

The cubic CEF Hamiltonian, choosing a fourfold axis as quantization axis, has the form~\cite{fulde:79}
\begin{equation}
	{\cal H}_\text{cubic}
	=
	B_4^{(4)}\left(O_4^0+5O_4^4\right)
	+B_6^{(4)}\left(O_6^0-21O_6^4\right),
	\label{eqn:cef:cubic}
\end{equation}
where the $O_l^m$ are functions of the angular-momentum operators called Stevens operator equivalents~\cite{hutchings:64} and the $B_l^{(4)}$ are numerical factors called crystal-field parameters. The eigenvalues are two doublets plus a quartet~\cite{Lea62}
\begin{align*}
	E_{\Gamma_6}
	&=
	840\left(B_4^{(4)}-30B_6^{(4)}\right),
	\\
	E_{\Gamma_7}
	&=
	-1080\left(B_4^{(4)}+14B_6^{(4)}\right),
	\\
	E_{\Gamma_8}
	&=
	120\left(B_4^{(4)}+168B_6^{(4)}\right)
\end{align*}
with the multiplets
\begin{align*}
	\Gamma_6
	&:\quad
	\frac12\sqrt{\frac53}\left|\frac72,\pm\frac72\right\rangle
	+\frac12\sqrt{\frac73}\left|\frac72,\mp\frac12\right\rangle,
	\nonumber\\
	\Gamma_7
	&:\quad
	\mp\frac{\sqrt3}2\left|\frac72,\pm\frac52\right\rangle
	\pm\frac12\left|\frac72,\mp\frac32\right\rangle,
	\nonumber\\
	\Gamma_8
	&:\quad
	\left\{\begin{aligned}
	&
	\mp\frac12\sqrt{\frac73}\left|\frac72,\pm\frac72\right\rangle
	\pm\frac12\sqrt{\frac53}\left|\frac72,\mp\frac12\right\rangle
	\\
	&
	\frac{\sqrt3}2\left|\frac72,\pm\frac32\right\rangle
	+\frac12\left|\frac72,\mp\frac52\right\rangle
	\end{aligned}\right..
	\nonumber
\end{align*}
Applying a tetragonal distortion to the cubic Hamiltonian in particular splits the $\Gamma_8$ quartet. Writing the crystal-field Hamiltonian as
\begin{align*}
	{\cal H}_\text{CEF}
	&=
	{\cal H}_\text{cubic}+\lambda{\cal H}_\text{tetragonal},
	\\
	{\cal H}_\text{tetragonal}
	&=
	\delta B_2^0O_2^0+\delta B_4^0O_4^0+\delta B_4^4O_4^4
	\\
	&\phantom{=}
	+\delta B_6^0O_6^0+\delta B_6^4O_6^4
\end{align*}
we find to lowest order in $\lambda$
\begin{align*}
	E_{\Gamma_6}
	&\to
	E_{\Gamma_6}+
	70\lambda\left[
	7\delta B_4^0+\delta B_4^4-15\left(3\delta B_6^0-\delta B_6^4\right)
	\right],
	\\
	E_{\Gamma_7}
	&\to
	E_{\Gamma_7}
	-90\lambda\left[
	7\delta B_4^0+\delta B_4^4+7\left(3\delta B_6^0-\delta B_6^4\right)
	\right],
	\\
	E_{\Gamma_8}
	&\to E_{\Gamma_8}
	\left\{
	\begin{aligned}
	&
	-6\lambda\left[
	\delta B_2^0
	+5\left(11\delta B_4^0-3\delta B_4^4\right)
	-105\left(11\delta B_6^0-\delta B_6^4\right)
	\right]
	\\
	&
	+6\lambda\left[
	\delta B_2^0
	+10\left(47\delta B_4^0-7\delta B_4^4\right)
	-210\left(9\delta B_6^0+5\delta B_6^4\right)
	\right]
	\end{aligned}
	\right..
\end{align*}
Interestingly the component $\delta B_2^0O_2^0$, absent in ${\cal H}_\text{cubic}$, only affects the $\Gamma_8$ states: $(O_2^0,O_2^2)$ transform like $\Gamma_3$. Group theory tells us that $\Gamma_3\otimes\Gamma_8=\Gamma_6+\Gamma_7+\Gamma_8$, therefore $\left\langle\Gamma_8\left|O_2^0\right|\Gamma_8\right\rangle\ne0$. In contrast $\Gamma_3\otimes\Gamma_{6,7}=\Gamma_8$, so $\left\langle\Gamma_6\left|O_2^0\right|\Gamma_6\right\rangle=\left\langle\Gamma_7\left|O_2^0\right|\Gamma_7\right\rangle=0$.

We expect $\left\langle\Gamma_8\left|O_2^0\right|\Gamma_8\right\rangle$ to be the largest contribution to ${\cal H}_\text{tetragonal}$ such that a description using the impact of $O_2^0$ only should be sufficient at least in case of small distortions, i.e,
\begin{align*}
	{\cal H}_\text{CEF}
	&\to
	{\cal H}_\text{cubic}+g_{zz}\epsilon_{zz}O_2^0,
	\\
	E_{\Gamma_8}
	&\to E_{\Gamma_8}
	\pm6g_{zz}\epsilon_{zz}
\end{align*}
with an applied strain $\epsilon_{zz}$ along one of the fourfold cubic axes and $g_{zz}$ the magneto-elastic coupling strength~\cite{Luethi05}.

\clearpage

\end{document}